\documentclass[twocolumn,pra,aps,showpacs,superscriptaddress]{revtex4-1}%
\usepackage{setspace}
\usepackage{graphicx,graphics}
\usepackage{amsmath}
\usepackage{amssymb}
\usepackage{epstopdf}
\usepackage{amstext}
\usepackage{amsthm}
\usepackage{amsfonts}
\usepackage{latexsym}
\usepackage{array}
\usepackage{xfrac}
\usepackage{color}
\usepackage{booktabs}
\usepackage{fancyhdr}
\usepackage{times}
\usepackage[utf8]{inputenc}
\usepackage[english]{babel}
\usepackage{color}
\usepackage{soul}

\begin{document}
	
	\title{Dark-dark-soliton dynamics in two 
density-coupled Bose-Einstein condensates} 
	
	\author{I. Morera}
	\affiliation{Departament de F\'isica Qu\`antica i Astrof\'isica, 
		Facultat de F\'{\i}sica, Universitat de Barcelona, Mart\'i i Franqu\`es 1, 08028 Barcelona, Spain}
	\author{A. Mu\~noz Mateo}
	\affiliation{Dodd-Walls Centre for Photonic and Quantum 
Technologies and Centre for Theoretical Chemistry and Physics, New Zealand 
Institute for Advanced Study, Massey University, Private Bag 102904 NSMC, 
Auckland 0745, New 
Zealand}
	\author{A. Polls}
	\affiliation{Departament de F\'isica Qu\`antica i Astrof\'isica, 
		Facultat de F\'{\i}sica, Universitat de Barcelona, Mart\'i i Franqu\`es 1, 08028 Barcelona, Spain}
	\affiliation{Institut de Ci\`encies del Cosmos (ICCUB), Universitat de Barcelona, Mart\'i i Franqu\`es 1, 08028 Barcelona, Spain}
	\author{B. Juli\'a-D\'iaz}
	\affiliation{Departament de F\'isica Qu\`antica i Astrof\'isica, 
		Facultat de F\'{\i}sica, Universitat de Barcelona, Mart\'i i Franqu\`es 1, 08028 Barcelona, Spain}
	\affiliation{Institut de Ci\`encies del Cosmos (ICCUB), Universitat de Barcelona, Mart\'i i Franqu\`es 1, 08028 Barcelona, Spain}
	\affiliation{Institut de Ci\`encies Fot\`oniques, Parc Mediterrani de la Tecnologia, 08860 Barcelona, Spain }

	\date{\today}
	
\begin{abstract}
We study the 1D dynamics of dark-dark solitons in the miscible regime of two 
density-coupled Bose-Einstein condensates having repulsive interparticle 
interactions  within each condensate ($g>0$). By using an 
adiabatic perturbation theory in the parameter $g_{12}/{g}$, we show that, 
contrary to the case of two solitons in scalar condensates, the interactions 
between solitons are attractive when the interparticle interactions between 
condensates are repulsive $g_{12}>0$. As a result, the relative motion of 
dark solitons with equal chemical potential $\mu$ is well approximated by 
harmonic oscillations of angular frequency 
$w_r=(\mu/\hbar)\sqrt{({8}/{15}){g_{12}}/{g}}$. 
We also show that in finite 
systems, the resonance of this anomalous excitation mode with the spin 
density mode of lowest energy gives rise to alternating dynamical instability 
and stability fringes as a function of the perturbative parameter.
In the presence of harmonic trapping (with angular frequency $\Omega$) the 
solitons are driven by the superposition of two harmonic motions at 
a frequency given by $w^2=(\Omega/\sqrt{2})^2+w_r^2$. When $g_{12}<0$, these 
two oscillators compete to give rise to an overall effective potential
that can be either single well or double well through a pitchfork bifurcation. 
All our theoretical results are 
compared with numerical solutions of the Gross-Pitaevskii equation for the 
dynamics and the Bogoliubov equations for the linear stability. A good 
agreement is found between them.
		
	\end{abstract}

	\maketitle

\section{Introduction}
Over the last two decades, Bose-Einstein condensation (BEC) has enabled the
study of numerous physical concepts \cite{Pitaevskii2003,Pethick2008}. Among 
them, an interesting scenario is the connection between the non-linear waves 
and atomic systems, that leads to the so-called matter-wave solitons 
\cite{Kevrekidis2008,Kevrekidis2015}. These structures emerge from the 
balance between linear dispersion and interactions, which are accounted for 
from non-linear terms in the equations of motion. Specifically, 
mean field descriptions of BECs, as 
provided by the Gross-Pitaevskii equation (GP), incorporate a 
non-linear term proportional to the interatomic interaction strength $g$, 
which is measured by the s-wave scattering length $a$. Depending on the sign 
and magnitude of the latter and the 
dimensionality of the system, a large number of structures can be found: 
dark \cite{Frantzeskakis2010} and bright \cite{Strecker2003} solitons, vortices 
\cite{Fetter2001}, etc.

Dark solitons in BECs are localized non-linear
excitations that present a notch in the condensate density and a phase 
step across its center. They have been observed in numerous 
experiments using different techniques 
\cite{Burger1999,Becker2008,Weller2008,Lamporesi2013,Anderson1999} and this has inspired a 
large number of theoretical works (see \cite{Frantzeskakis2010} and references 
therein). Specifically, the problem of dark solitons in BECs confined by 
parabolic external traps has been extensively studied \cite{William1997,Muryshev1999,Busch2000,Pelinovsky2005}. 

Another interesting aspect of BECs is the study of multi-component systems,
which can be described by a set of coupled GP equations. These systems were soon 
realized in ultracold-gas experiments by coupling two condensates made of 
either different atomic 
species \cite{Cornell1998} or different
hyperfine states of the same atomic species \cite{Myatt1997}. This fact 
inspired the study of matter-wave solitons in these settings, and  new families 
of solitonic structures have been found: dark-dark 
\cite{Ohberg2001,Hoefer2011,Yan2012}, dark-bright 
\cite{Middelkamp2011,Yan2011,Achilleos2011}, dark-antidark \cite{Danaila2016}, etc. 
Particular attention has been given to the so-called Manakov limit in 1D 
settings, where the intra- and inter-condensate particle interactions match 
\cite{Yan2012}.

The aim of this work is to study the dynamics of dark-dark solitons in a one-dimensional setting of two
density-coupled condensates out of the Manakov limit, for varying coupling 
between condensates. 
Specifically, we consider a mean-field description of BECs composed by two 
hyperfine states of the same alkali species, and inspect both the case without any 
axial trap, termed untrapped, and the case with a harmonic trap in the axial direction. 
Within the Hamiltonian approach of the perturbation
theory for solitons \cite{Uzunov1993,Kivshar1994,Frantzeskakis2002}, we obtain 
analytical expressions for the adiabatic evolution of the dark-dark soliton.
For the untrapped case we find that when the interaction between components 
is repulsive the dark-dark soliton can be seen as a bound state of two dark 
solitons performing a relative harmonic motion. For attractive interactions 
such a bound state can not exist because the dark solitons repel each other. An 
equivalent study is also performed for the confined case, where the competition 
between soliton interactions and harmonic trapping 
is found to produce sizable changes on the dynamics, leading for 
example to bound states even with attractive interactions. Our 
analytical results are tested with different numerical techniques:
direct simulations of the Gross-Pitaevskii equations, Fourier 
analysis to extract characteristic frequencies of the motion, and numerical 
solution of  the Bogoliubov equations for the linear excitations of 
stationary states. In what follows, we first
introduce the theoretical model in section II, and next we 
show our numerical results in section III. To sum up, we present our 
conclusions in section IV.

\section{Theoretical model}
The dynamics of a BEC  at 
zero temperature can be accurately described within the 
mean field approach in terms of 
a wave-function $\psi(\mathbf{r},t)$. In a one dimensional setting, the 
wave functions of two trapped, density-coupled BECs are 
governed by corresponding Gross-Pitaevskii equations
\begin{align}
\begin{aligned}
&i\hbar \frac{\partial}{\partial t} \psi_1 = 
\left(\frac{-\hbar^2}{2m}\frac{\partial^2}{\partial z^2}+\frac{m\Omega_z^2 
z^2}{2}+ g |\psi_{1}|^2 + g_{12} |\psi_{2}|^2 \right)\psi_{1}
\\
&i\hbar \frac{\partial}{\partial t} \psi_2 = 
\left(\frac{-\hbar^2}{2m}\frac{\partial^2}{\partial z^2}+\frac{m\Omega_z^2 
z^2}{2}+ g |\psi_{2}|^2 + g_{12} |\psi_{1}|^2 \right)\psi_{2}\, ,
\end{aligned}
\label{eq:ebmr_dim}
\end{align} 
where $\Omega_z$ is the angular frequency of the axial harmonic trapping, 
$g=2\hbar \Omega_\perp a$ is the reduced 1D strength of the interaction between 
particles of the same condensate, which is proportional to the energy of the 
transverse trapping  $\hbar \Omega_\perp$ and to the scattering length $a$, and 
$g_{12}=2\hbar \Omega_\perp 
a_{12}$ is the strength of the density coupling between condensates, 
proportional to the scattering length between particles of different 
condensates $a_{12}$.

A stationary soliton solution to Eq.~(\ref{eq:ebmr_dim}) $\psi_i(z,t)$, 
with $i=1,2$, can be written as the product of a time-independent (and real 
function) background component $\phi_i$ and the soliton excitation $v_i$, so 
that
\begin{align}
\psi_i(z,t)=\phi_i(z) v_i(z,t) e^{-i \mu_i t/\hbar},
\label{DS}
\end{align}
where $\mu_i$ is the chemical potential. It is important to remark that both 
$\psi_i$ and $\phi_i$ satisfy the coupled GP equations for the same 
chemical potential:   
\begin{align}
&\mu_i \phi_i = 
\left(-\frac{1}{2}\frac{\partial^2}{\partial z^2}+\frac{1}{2}\Omega^2 z^2+
g \phi_{i}^2 + g_{12} \phi_{j}^2 \right)\phi_{i}
\label{eq:ebmr20}
\\
&i \frac{\partial}{\partial t} \psi_i = 
\left(-\frac{1}{2}\frac{\partial^2}{\partial z^2}+\frac{1}{2}\Omega^2 z^2+
g |\psi_{i}|^2 + g_{12} |\psi_{j}|^2 \right)\psi_{i} \,,
\label{eq:ebmr2}
\end{align} 
where $\Omega=\Omega_z/\Omega_\perp$ is the trap aspect ratio, and we have 
written the equations in dimensionless form by using transverse trap units, 
$a_\perp=\sqrt{\hbar/m\Omega_\perp}$ as unit length and 
$t_\perp=\Omega_\perp^{-1}$ as unit time, and for the sake of a 
simple notation have kept the same letters for the dimensionless couplings 
${g}$ and ${g}_{12}$. So, from 
Eqs.~(\ref{eq:ebmr20})-(\ref{eq:ebmr2}) we 
can obtain corresponding equations for the soliton wave functions $v_i(z,t)$
\begin{align}
&\left(i \frac{\partial}{\partial t}  
+ \frac{1}{2}\frac{\partial^2 }{\partial z^2}- g\phi^2_{i} 
\left(|v_{i}|^2-1\right) - g_{12}\phi^2_j\left(|v_{j}|^2-1\right)\right)v_i=
\nonumber\\
&\hspace{.5cm} =-\frac{\partial v_i}{\partial z}\,\frac{\partial \ln 
\phi_i}{\partial z}\,,
\label{eq:v}
\end{align}
where the external trap does not appear explicitly, although its
information is encoded in the background wavefunctions $\phi_i$.

\subsection{Untrapped case}
Here we set $\Omega=0$, hence the ground state solutions to Eq. 
(\ref{eq:ebmr20}) are the constant density $|\phi_i|^2=n_i$ states
\begin{equation}
n_i=\frac{g\mu_i-g_{12}\mu_j}{g^2-g^2_{12}}. 
\label{eq:n_i}
\end{equation}
The equations of motion for soliton states Eq. (\ref{eq:v}) can be 
written as
\begin{align}
\left(i \frac{\partial }{\partial t}  +
\frac{1}{2}\frac{\partial^2 }{\partial z^2}
- g\,n_i\, \left(|v_{i}|^2-1\right)
- g_{12}\,n_j\,  \left(|v_{j}|^2-1\right)\right)v_i=0.
\label{eq:v2}
\end{align}

As we said before, we are going to deal with the coupling between condensates 
in a perturbative way. Then, assuming that $g_{12}<<g$, the background 
densities Eq. (\ref{eq:n_i}) become $n_i\approx \mu_i/g-(g_{12}/g)\,\mu_j/g$, 
and neglecting terms proportional to $g_{12}^2$ in Eq. (\ref{eq:v2}) we get
\begin{align}
\begin{aligned}
&i \frac{\partial v_i}{\partial t}  + 
\frac{1}{2}\frac{\partial^2 v_i}{\partial z^2}- \mu_i  
\left(|v_{i}|^2-1\right)v_i = 
\\
&=- \frac{g_{12}}{g}\mu_j 
\left(|v_{i}|^2-|v_{j}|^2\right)v_i\equiv \frac{g_{12}}{g}\mu_j 
P(v_i,v_j).
\end{aligned}
\label{eq:v3}
\end{align}
On the left hand side of this equality we have the well known non-linear 
Schr\"odinger equation for the wave function $v_i$, whereas on the right hand 
side we get a perturbative term in the parameter $g_{12}/g$.
In order 
to obtain analytical expressions for the interactions between solitons, 
from now on we will focus on the case of equal 
backgrounds $\phi_1=\phi_2=\phi$, and then $\mu_1=\mu_2=\mu=(g+g_{12})n$. So that, by dividing the whole equation \eqref{eq:v3} by $\mu$, we set new units for space $\hbar/\sqrt{m\mu}$ and time $\hbar/\mu$. In these units we introduce the general dark soliton solution
\begin{align}
\begin{aligned}
&v_i(z,t)=\cos\varphi_i \tanh \zeta_i + i \sin\varphi_i,
\end{aligned}
\label{soliton}
\end{align}
where $\zeta_i=\cos\varphi_i(z-t\sin\varphi_i )$, and $\varphi_i$ 
parametrizes the soliton darkness, that is $|v_i|^2=(1-\cos^2\varphi_i \,  
\mbox{sech}^2 \, \zeta_i)$, and the soliton velocity $\dot z_0=\sin \varphi_i$, 
that is the time derivative of the soliton position $z_0$, taken at minimum 
density.

In the absence of perturbation, $P(v_i,v_j)=0$, the solution Eq.
(\ref{soliton}) describes overlapping solitons moving at constant velocity, and 
hence $\ddot z_0=\dot\varphi_i=0$. However, the effect of the perturbation 
$P(v_i,v_j)$ leads to the adiabatic time evolution of the soliton parameters, 
$(\varphi, z_0)\rightarrow (\varphi(t), z_0(t))$. This evolution can be 
described by the equation \cite{Kivshar1994} 
\begin{align}
\dot\varphi_i=\frac{g_{12}/g}{2 \cos^2 \varphi_i \sin \varphi_i}
\Re\left[\int_{-\infty}^{\infty} P(v_i,v_j)\frac{\partial v_i^*}{\partial t} 
dz\right].
\label{pert}
\end{align}

\begin{figure}[!tb]
	\centering
	\includegraphics[width=\columnwidth]{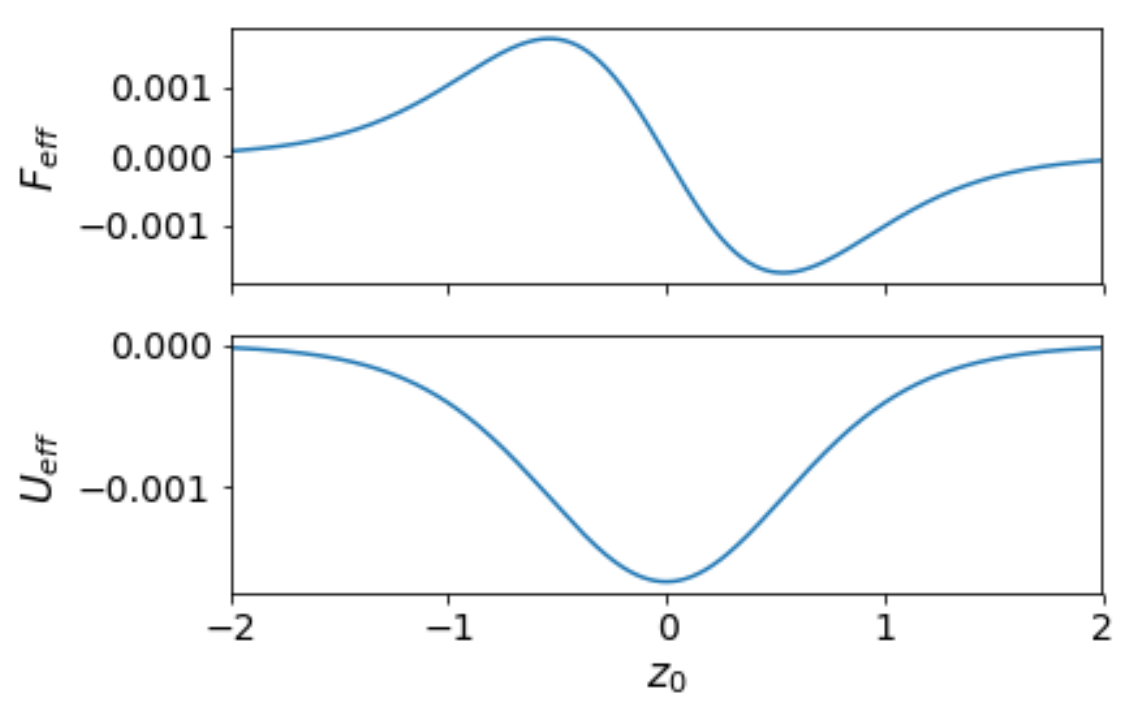}
	\caption{Top panel: Effective force experienced by a dark soliton 
forming a dark-dark soliton system for $g_{12}/g=0.01$ in the absence of 
trapping. Bottom panel: 
Soliton interaction potential Eq. (\ref{eq:int_pot}) from which the above
effective force is derived. The solitons oscillate in a relative motion 
around the minimum at $z_0=0$.}
	\label{fig:pert}
\end{figure}

Assuming solitons with high density depletions $\varphi_i<<1$ and separated 
by a relative distance $2z_0$, such that $\zeta_1=\cos \varphi_1 
\left(z-z_0\right)$ and $\zeta_2=\cos \varphi_2 \left(z+z_0\right)$, Eq. 
(\ref{pert}) gives
\begin{align}
\dot\varphi_1=-\frac{g_{12}}{g}f(z_0)\equiv \, F_{\rm eff}^{(i)}(z_0),
\label{eq:phi}
\end{align}
where 
$f(z_0)=8\sinh(2z_0)e^{6z_0}(3-3e^{8z_0}+(1+4e^{4z_0}+e^{8z_0})4z_0)/(e^{4z_0}
-1)^5$. As 
can be seen, for $z_0=0$, that is for overlapped solitons, $\dot \varphi_1=0$ 
and the solitons evolve without relative motion.
Since, in this limit ($\varphi<<1$) $\dot z_0\approx 
\varphi$, Eq. (\ref{eq:phi}) is also an equation for $\ddot z_0$, with the 
right hand side playing the role of a classical force 
$F_{\rm eff}^{(i)}(z_0)$   derived from an effective 
potential $U_{\rm eff}^{(i)}(z_0)$ :
\begin{align}
\ddot z_0=-\frac{d U_{\rm eff}^{(i)}(z_0)}{dz_0}.
\label{eq:rel}
\end{align}
This potential 
accounts for the interactions between the solitons and is the source of their 
relative motion. After integration of the force Eq. (\ref{eq:phi}) over $z_0$, 
we get 
\begin{align}
U_{\rm eff}^{(i)}(z_0)=\frac{g_{12}}{g}\,\frac{2(1+e^{4z_0})(1-2z_0)-4}
{ e^{-4z_0}\,(e^{4z_0}-1)^3}.
\label{eq:int_pot}
\end{align}
As it has been anticipated, it presents a minimum at $z_0=0$, allowing for 
the existence of a bound state made of overlapped solitons, and tends 
exponentially to zero for finite values of the inter-soliton distance $2z_0$ 
(see Fig. \ref{fig:pert}). For small separation $z_0\rightarrow 0$, the 
Taylor expansion 
reads
\begin{align}
U_{\rm eff}^{(i)}(z_0)\sim 
\frac{g_{12}}{g}\left(-\frac{1}{6}+\frac{4}{15} z^2_0 \right) +O[z^4_0]\,,
%-\frac{16}{63}\frac{g_{12}}{g} z^4_0 +O[z^6_0].
\end{align}
and the equation of relative motion Eq.~(\ref{eq:rel}) simplifies to that 
of a harmonic oscillator, $\ddot z_0+w_r^2 \,z_0=0$, with angular frequency 
(for generic chemical potential $\mu$)
\begin{equation}
 w_r=\mu\sqrt{\frac{8}{15}\frac{g_{12}}{g}}.
 \label{eq:freq_rel}
\end{equation}
This is the frequency of small oscillations of the two solitons around 
their center of mass, moving at constant velocity $v_{CM}=(\sin 
\varphi_1+\sin \varphi_2)/2$. 
It is also interesting to note that this model predicts an instability 
(imaginary frequency) when $g_{12}<0$, due to the fact that the effective 
potential $U_{\rm eff}^{(i)}$ presents a maximum for attractive interparticle 
interactions between particles of different condensates, and as a consequence, 
it prevents the existence of a bound solitonic state in this case.

\subsection{Trapped case}
In the presence of axial harmonic trapping $V(z)=\frac{1}{2}\Omega^2 z^2$, 
we rely on the Thomas-Fermi (TF) approximation to study the strong 
(interparticle) interacting 
regime ($\mu_i>>\Omega$) \cite{Pitaevskii2003}. There the 
inhomogeneous ground state densities, $n_i(z)=|\phi_i|^2$, are given by
\begin{align}
n_i(z)=\frac{g\mu_i(z)-g_{12}\mu_j(z)}{g^2-g^2_{12}},
\label{eq:TF}
\end{align}  
where $\mu_i(z)=\mu_i-V(z)$ are the local chemical potentials.
In the limit of $g_{12}<<g$, we get
\begin{align}
g\,n_i(z)={\mu_i(z)}-\frac{g_{12}}{g}{\mu_j(z)},
\end{align}
and, as a result,  the equations of motion for soliton solutions  
Eq.~(\ref{eq:v}) become 
\begin{align}
&i \frac{\partial v_i}{\partial t}  + 
\frac{1}{2}\frac{\partial^2 v_i}{\partial z^2}- \mu_i(z)  
\left(|v_{i}|^2-1\right)v_i = 
\nonumber \\
&=- \frac{g_{12}}{g}\mu_j(z) \left(|v_{i}|^2-|v_{j}|^2\right)v_i 
-\frac{\partial v_i}{\partial z}\,\frac{d\ln \phi_i}{dz} .
\end{align}
The local chemical potentials and the last term in the right hand side 
of this equation are the main differences with respect to the untrapped case 
Eq. (\ref{eq:v3}).
In spite of these differences, and due to the fact that the background 
densities change slowly inside the TF regime ($dV/dz\rightarrow 0$), an 
analogue perturbative approach can still be followed, and the
governing equation for the dark soliton $v_i$ is
\begin{align}
i \frac{\partial v_i}{\partial t}  + \frac{1}{2}\frac{\partial^2 v_i}{\partial 
z^2}-\mu_i(z) v_i \left(|v_{i}|^2-1\right)=P(v_i,v_j;V(z)),
\label{eq:v5}
\end{align}
where the perturbation $P(v_i,v_j;V)$ is given by
\begin{align}
&P(v_i,v_j;V)\approx - \frac{g_{12}}{g} \mu_j(z)
\left(|v_{i}|^2-|v_{j}|^2\right)v_i+
\nonumber\\
&\hspace{.5cm} 
+\frac{1}{2\mu(z)}\frac{dV}{dz}\frac{\partial v_i}{\partial 
z},
\end{align}
which, along with the term associated to the interaction between dark 
solitons, contains a new term introduced by the external trap \cite{Frantzeskakis2010}.

Again, we focus on the symmetric case $\mu_1=\mu_2=\mu$, where  
the local chemical potential reads $\mu(z)=\mu-V(z)=(g+g_{12})\,n(z)$. In the particular case of motion around the center of the harmonic potential
$\mu(z)\rightarrow \mu$, and by following the same procedure as in the untrapped case
(with $\hbar/\sqrt{m\mu}$ and $\hbar/\mu$ as space and time units respectively) we can obtain a particle-like evolution 
for the soliton position $z_0$

\begin{figure}[!t]
	\centering
	\includegraphics[width=\linewidth]{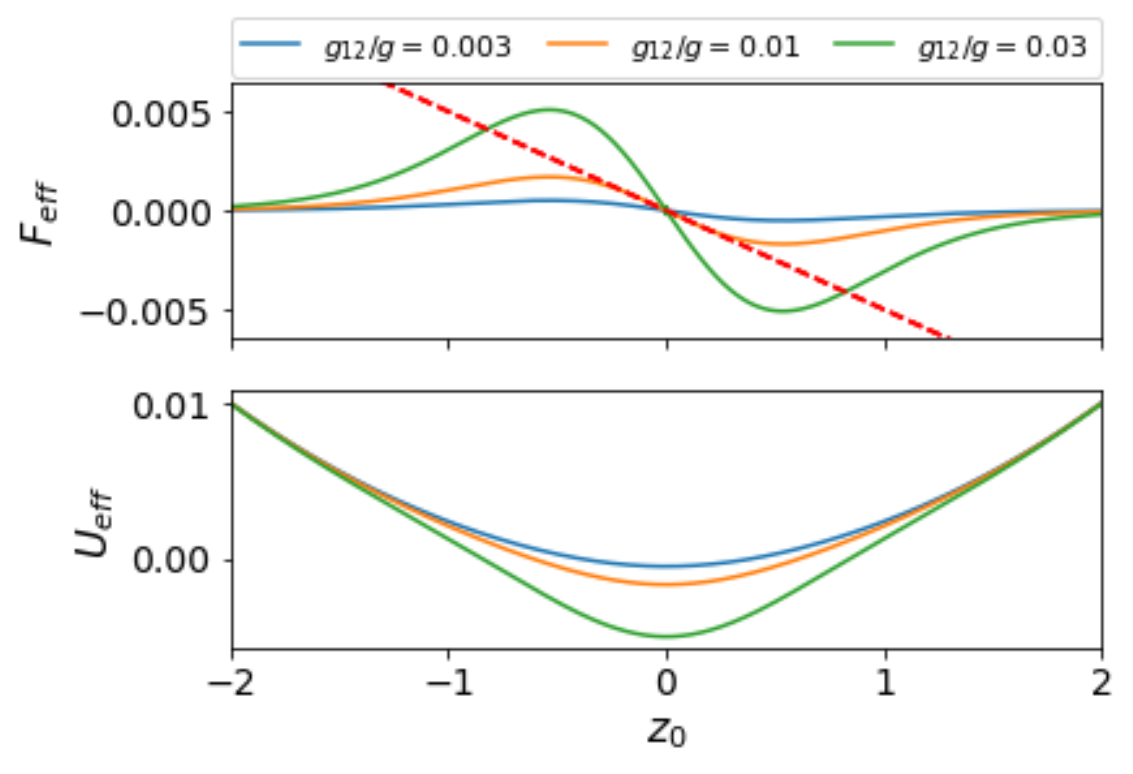}
	\caption{(Color online)  
	 Forces (top panel) and total, single-well potential (bottom panel) 
experienced by each dark soliton forming a dark-dark soliton system with 
$\mu=1$ and $g_{12}>0$ in the presence of a harmonic trap with $\Omega=0.1$. 
The dashed (red) line represents the force caused by the trap, whereas the 
continuous lines represent the effective force due to the interaction with the 
other dark soliton.}
	\label{fig:potg12posa001}
\end{figure}

\begin{align}
\ddot{z}_0=-\frac{\Omega^2}{2 \mu^2} z_0- 
F_{\rm eff}^{(i)} ,
\label{eq:freq_trap}
\end{align}
where the right hand side is the total effective force acting on the 
soliton $F_{\rm eff}=-dU_{\rm eff}/d\,z_0$, and embraces the superposition of the 
action of the external trap and the soliton interaction. 
For small-amplitude oscillations $z_0\approx 0$, we get
\begin{align}
\ddot{z}_0=-\left(\frac{\Omega^2}{2\mu^2}+\frac{8}{15}\frac{g_{12}}{g}
\right)z_0+\frac{64}{63}\frac{g_{12}}{g}z_0^3+O(z_0^5),
\label{eq:cl2}
\end{align}
where the balance between the two 
harmonic force terms (inside the parenthesis) leads to different scenarios 
depending on the sign of the coupling interaction $g_{12}$.

\begin{figure}[!t]
	\centering
	\includegraphics[width=\columnwidth]{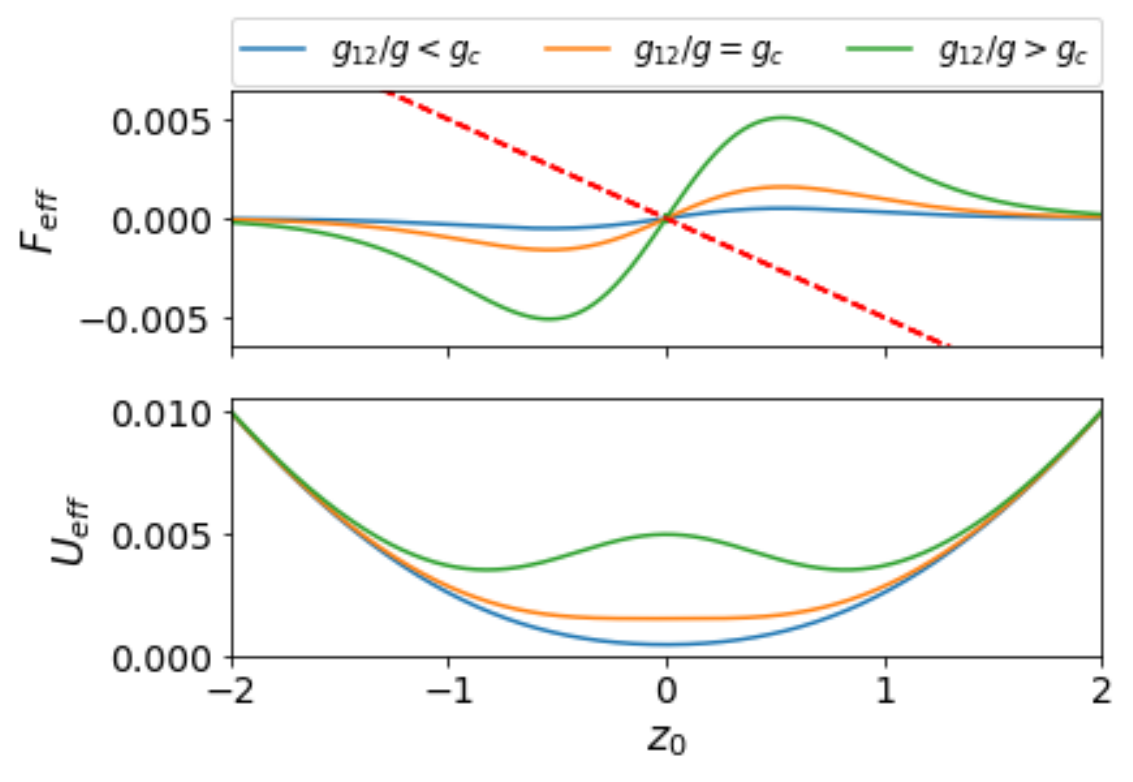}
	\caption{(Color online) Same as Fig. \ref{fig:potg12posa001} for  
$g_{12}<0$.  Three cases are depicted for different $g_{12}/g$ values, below 
(blue), at (orange) and above (green) the threshold $g_c$ indicating the 
change in the shape of the total effective potential from single well 
$g_{12}/g< c_c$ to double well $g_{12}/g>g_c$ (see text).  }
	\label{fig:potg12nega001}
\end{figure}

\subsubsection{Repulsive coupling: ${g_{12}>0}$}
Figure \ref{fig:potg12posa001} shows the effective force and effective 
potential defined in Eq. (\ref{eq:freq_trap}) for varying values of 
$g_{12}/g$ at a given trapping $\Omega=0.1$.
As can be seen, the different parameters do not produce qualitative changes in 
the potential, which presents a single minimum capable to bound the 
coupled solitons. As in the untrapped case, our theory predicts that the dark 
solitons will perform small-amplitude oscillations around $z_0=0$
with angular frequency (in transverse oscillator units)
\begin{equation}
w=\left(\frac{\Omega^2}{2}+\frac{8}{15}\frac{g_{12}}{g} 
\mu^2\right)^\frac { 1 } { 2 }.
\label{eq:fre} 
\end{equation}

As we will see later, our numerical results demonstrate that the minimum 
supports a bound state that exists and is stable for small $g_{12}/g$. However, 
for increasing values of this parameter both stable and unstable 
cases can be found depending on the particular valued of the chemical potential 
of the system.

\subsubsection{Attractive coupling: ${g_{12}<0}$}

The phenomenology is richer for attractive interactions between 
particles of different condensates. In contrast to the untrapped case, the 
presence of the 
harmonic potential allows for the generation of local minima in the total 
effective potential that can support stationary states made of
two solitons. As shown in Fig. \ref{fig:potg12nega001}, by 
increasing the parameter  $|g_{12}/g|$, the total effective potential modifies 
from a single well (an also a unique fixed point $z_0^*=0$ in the equation of 
motion) to  a double well potential (with three fixed points).  
Specifically, the system shows a pitchfork bifurcation at 
$g_c=|g_{12}/g|={15}(\Omega/\mu)^2/16$, hence for $|g_{12}/g|>g_c$ 
the fixed 
point at $z_0^*=0$ loses its stability and two new off-center, stable 
fixed points appear. From  Eq. (\ref{eq:cl2}) we get their position at 
\begin{equation}
z_{0_\pm}^{*} = \pm \frac{1}{\sqrt{\mu}}\sqrt{\frac{21}{40}-
\frac{63}{128} \left|\frac{g}{g_{12}}\right| 
\left(\frac{\Omega}{\mu}\right)^2 }.
\label{eq:fixed_double}
\end{equation}

\begin{figure}[!t]
\centering
\includegraphics[width=\columnwidth]{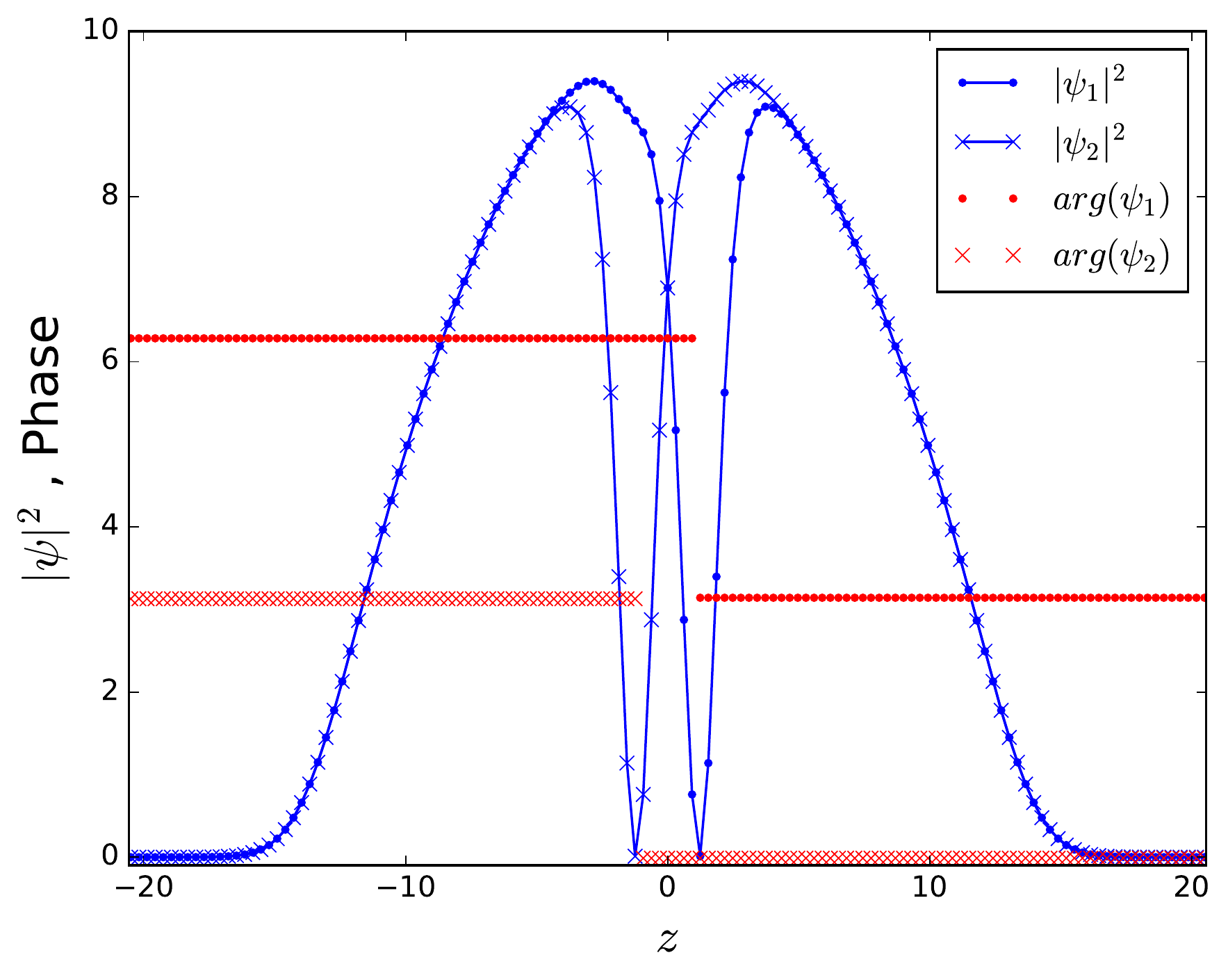}
\caption{(Color online) Stationary state made of two solitons situated at the 
fixed points $z_{0_\pm}^*$ of the effective potential (see text) for 
$\mu=10\,\Omega$ and $g_{12}/g=-0.1$.}
\label{fig:separated}
\end{figure}
 As can be noted, the separation between minima increases with 
$\mu/\, \Omega$ for given $g_{12}/g$, and saturates at a distance of 
$2z_0^*=1.44$. 
Fig. \ref{fig:separated} shows an example of a stationary state with 
two separated solitons occupying the two minima of the effective potential at 
$\mu=10 \, \Omega$ and $g_{12}/g=-0.1<g_c$.

For small distances $\epsilon(t)$ around the fixed 
points $z_{0_\pm}^*$, from the substitution of $z_0(t)=z_0^*+\epsilon(t)$ in 
Eq. (\ref{eq:cl2}), the solitons oscillate according to
\begin{equation}
\ddot{\epsilon}=-\left( w_1(z_{0_\pm}^*)^2 +w^2 \right) \epsilon \equiv 
-w_{\rm eff}^2 \epsilon ,
\label{eq:epsilon} 
\end{equation}
up to linear terms in the perturbation $\epsilon(t)$, where $w$ is the 
angular frequency given by \eqref{eq:fre}, and $ w_1(z_0)^2=-2 f(z_0)  
\left(  \mbox{sech}(z_0)- {5}/{\sinh(2z_0)} + 3\right)-$ 
$\left( \left(1+4e^{4z_0} +e^{ 8z_0}\right)+
\left(-12e^{8z_0}+\left(16e^{4z_0}+8e^{8z_0}\right) z_0 \right) \right)$
$ 64e^{6z_0} / \left(e^{4z_0}-1\right)^5 $.

\begin{figure}[!t]
	\centering
	\includegraphics[width=\columnwidth]{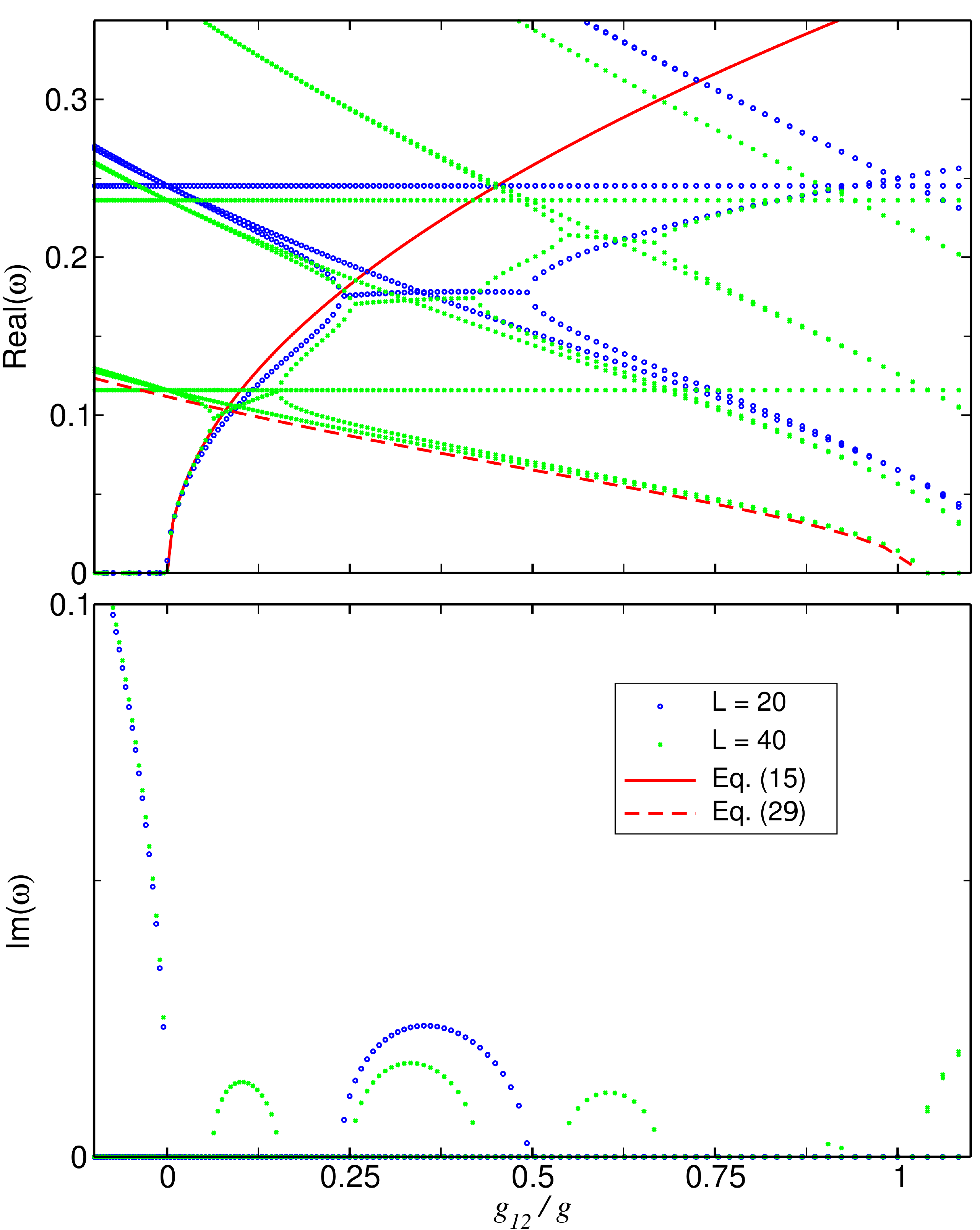}
	\caption{(Color online) Bogoliubov modes in 1D rings of 
		different size (untrapped case). The upper panel depicts the real part of the 
		frequencies and the lower panel the imaginary parts. The solid and dashed lines 
		correspond to Eqs. (\ref{eq:freq_rel}) and (\ref{eq:spin}), respectively, the 
		crossing of which indicates the appearance of instabilities.}
	\label{fig:bog_notrap}
\end{figure}

\section{Numerical results}
In what follows, in order to test our analytical predictions on the 
dark-dark-soliton dynamics, we first numerically solve the GP equation to 
obtain these stationary states for varying chemical potentials 
and interactions strengths. Afterwards, the soliton stability is monitored 
both in the nonlinear regime (by simulating the real time evolution with the GP 
equation) and by linear analysis around the equilibrium states. The latter is 
performed by solving the Bogoliubov equations  for the linear 
excitations $[u(z,t),v(z,t)]$ around the dark-dark soliton states 
$\psi(z,t)=\exp(-i \mu t)[\psi(z) +\sum_\omega(u \,e^{-i\omega t}+ v^* 
e^{i\omega t})]$. These equations are:

\begin{align}
B 
\left(\begin{array}{c}
u_1 \\
v_1 \\
u_2 \\
v_2
\end{array}\right)
= \omega 
\left(\begin{array}{c}
u_1 \\
v_1 \\
u_2 \\
v_2
\end{array}\right),
\label{eq:BdG}
\end{align}
where:
\begin{align}
B= 
\left(\begin{array}{cccc}
h_1 & g\psi_1^2 & g_{12}\psi_ 1 \psi_2^* & g_{12}\psi_1 \psi_2 \\
-g\psi_1^{*2} & -h_1 &  -g_{12}\psi_1^* \psi_2^* &  -g_{12}\psi_1^* \psi_2 \\
g_{12}\psi_1^* \psi_2 &  g_{12}\psi_1 \psi_2 & h_2 &  g\psi_2^2 \\
-g_{12}\psi_1^* \psi_2^* & - g_{12}\psi_1 \psi_2^* &-  g\psi_2^{*2} & -h_2
\end{array}\right).
\end{align}
Here $h_i=-\frac{1}{2} \frac{\partial^2}{\partial z^2} + \frac{1}{2}\Omega^2z^2 + 2g |\psi_i|^2 + g_{12} |\psi_{j\neq i}|^2 - \mu_i$.
They can be seen as an eigenvalue problem with a non-trivial solution given by 
${\rm det}|B- \omega I|=0$.

Apart from the perturbative case $g_{12}<<g$, we also explore numerically the 
stability of overlapped solitons in the whole miscible regime 
$g_{12}<g$, and show that the finite size of the system determines the 
stability properties.

\subsection{Untrapped case}
\begin{figure}[!b]
	\centering
	\includegraphics[width=\columnwidth]{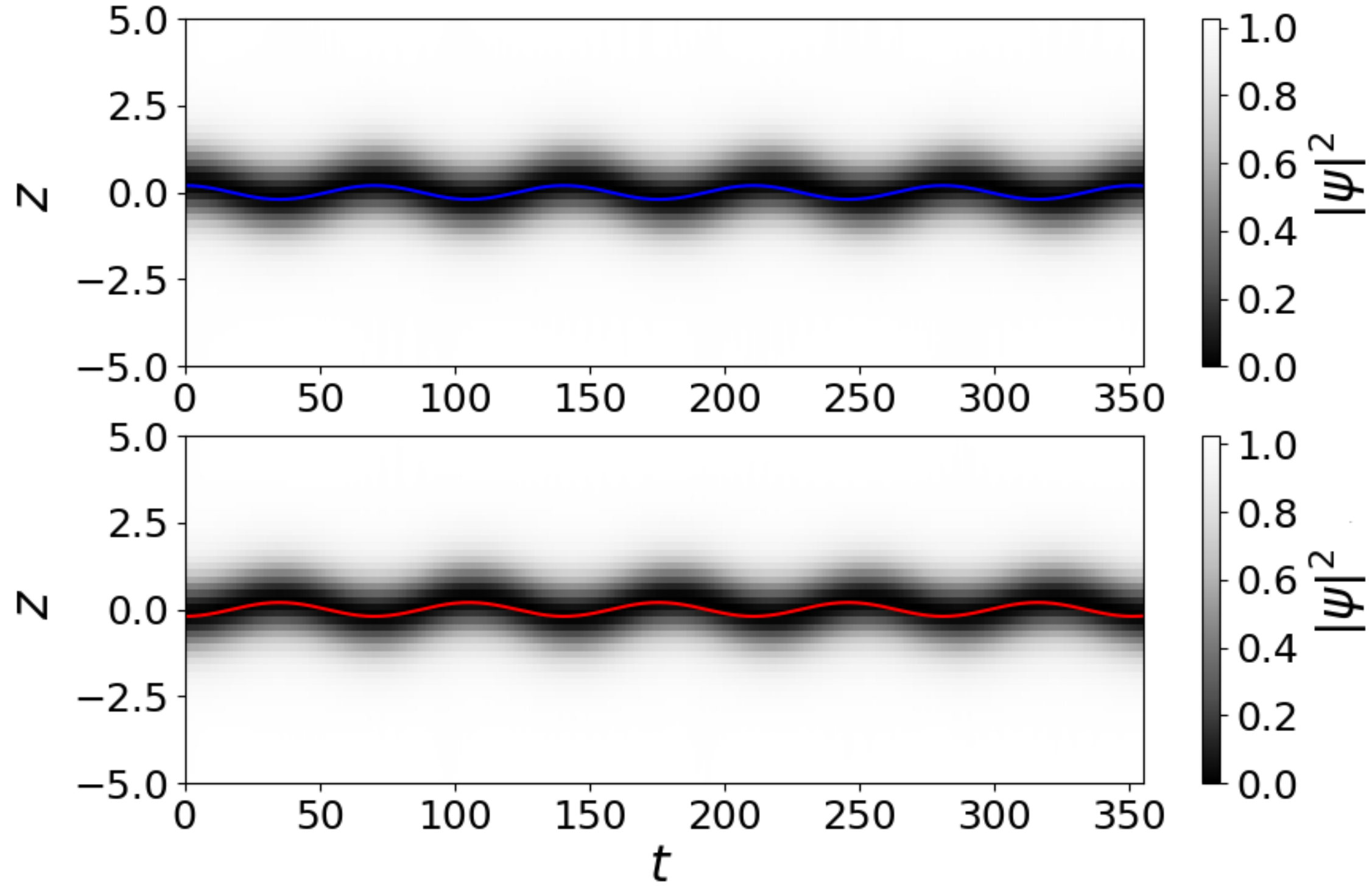}
	\caption{(Color online) Spatio-temporal evolution of the density of 
		two untrapped dark-dark solitons with 
$\mu=1$, $g_{12}=0.015$ and $g=1$ obtained 
		from the numerical solution of the GP Eq. \eqref{eq:ebmr_dim}.
		The top and down panel correspond to component 1 and 2 
respectively. The continuous (colour) lines represent the evolution of the dark 
solitons positions given by Eq. (\ref{eq:freq_rel}).}
	\label{fig:osci003}
\end{figure}
\begin{figure}[!t]
	\centering
	\includegraphics[width=\columnwidth]{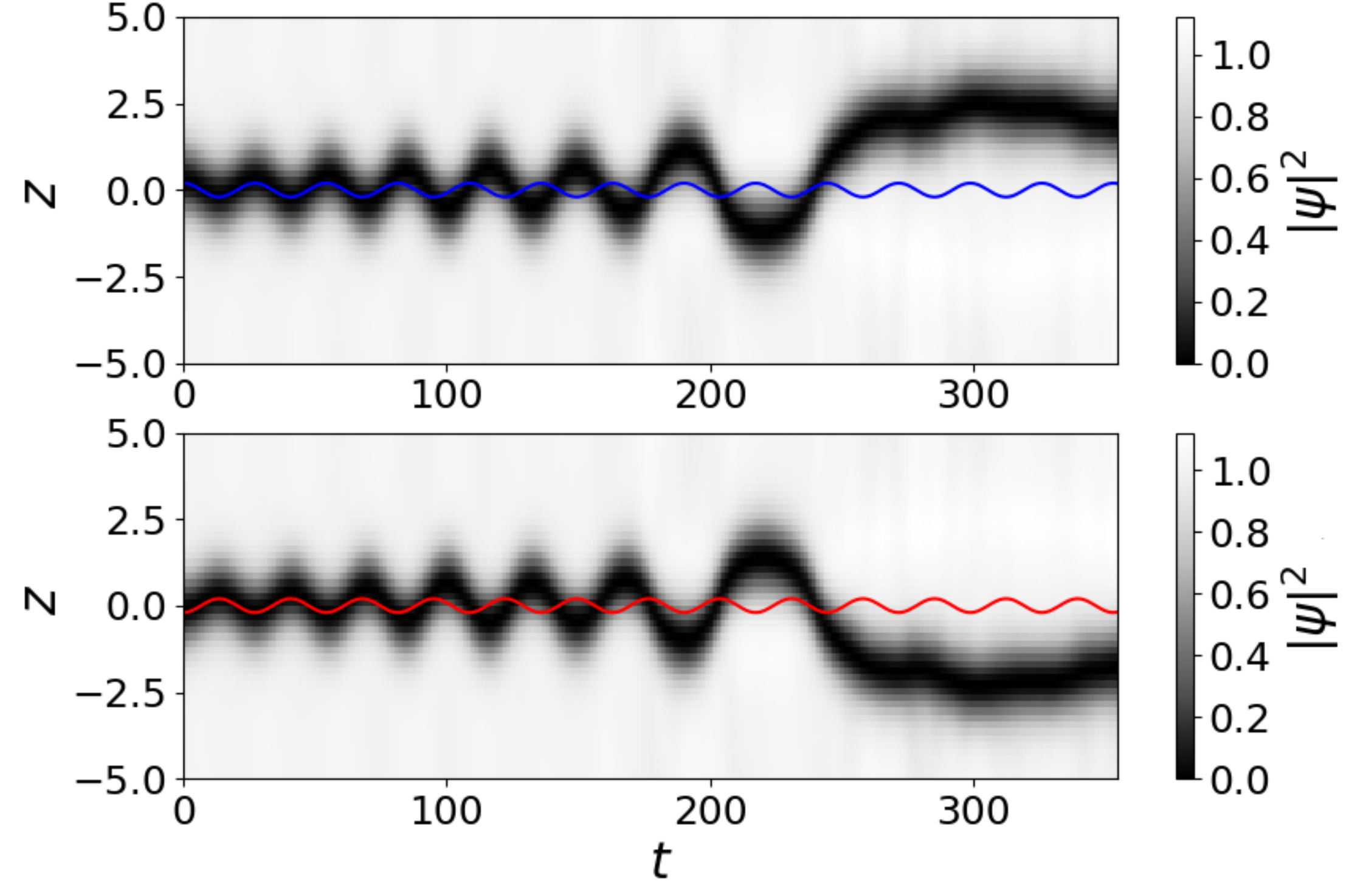}
	\caption{Same as Fig. \ref{fig:osci003} but with $g_{12}=0.1$. }
	\label{fig:osci01}
\end{figure}
\begin{figure}[!b]
	\centering
	\includegraphics[width=\columnwidth]{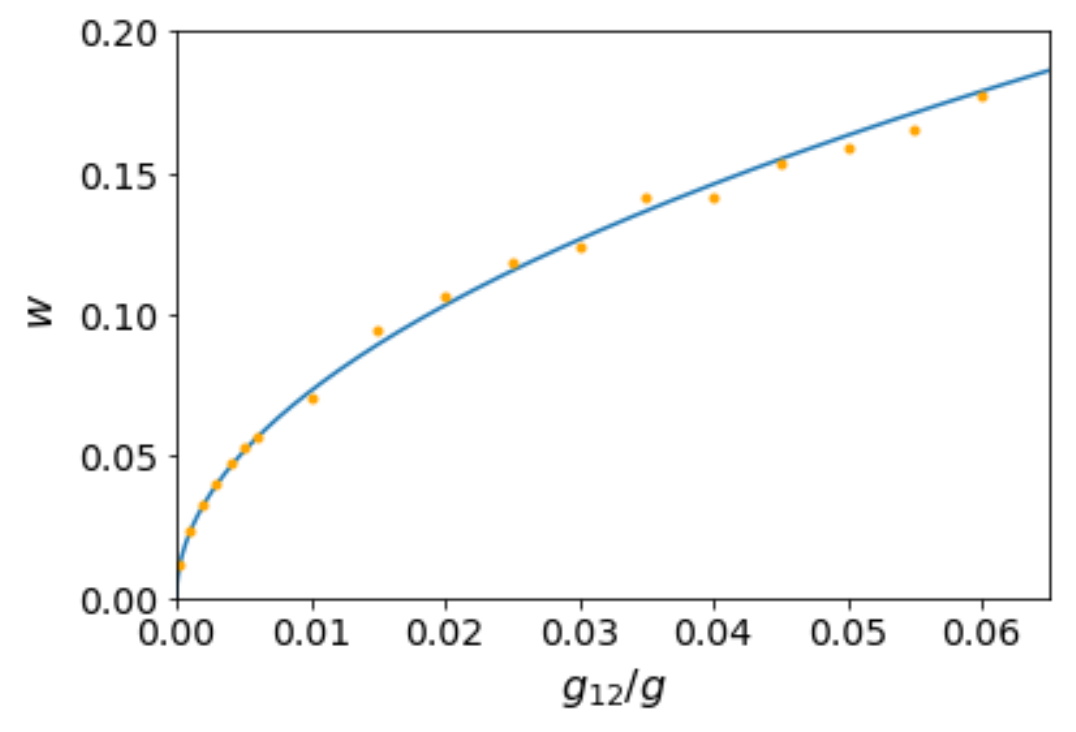}
	\caption{(Color online) Soliton relative motion frequency as a function 
 of the interparticle strength ratio in the absence of external trap. The 
continuous (blue) line
 represents our theoretical prediction Eq. (\ref{eq:freq_rel}) and the 
(orange) dots are values	extracted by means of Fourier analysis from the
numerical solution of the time dependent GP equation.}
	\label{fig:fourier}
\end{figure}
The excitation frequencies of the stationary states can be readily 
extracted by solving the Bogoliubov equations Eq. (\ref{eq:BdG}). In 
the Appendix we analytically show that there are two zero modes of excitation 
at $g_{12}/g=0$, and $1$, and hence that the system of overlapped dark solitons 
is expected to be unstable in between. However such analysis assume an infinite 
system, and the situation is quite different in systems of finite size, where 
ranges of dynamical stability can be found. Fig. \ref{fig:bog_notrap} shows two 
examples of this phenomenon, where the Bogoliubov modes are computed for 
overlapped solitons with the same chemical potential 
$\mu=0.5$ in 1D rings of different sizes $L=20,\; 40$.
The instability emerge from a Hopf bifurcation, which occurs due to the 
collision of two excitation modes (see the discussion about this collision in 
the next section): one mode associated to the 
oscillations around the minimum of the effective potential of 
soliton interactions, given by expression (\ref{eq:freq_rel}) and represented 
in Fig. \ref{fig:bog_notrap} by the solid red curve, and the background 
spin density mode of lowest energy (represented by the dashed curve for 
the longer ring), which is given by the analytical expression (in full units) 
\cite{Abad2013}:
\begin{equation}
\hbar\omega=\sqrt{\frac{\hbar^2 k^2}{2m} \left( \frac{\hbar^2 k^2}{2m} + 
2(g-g_{12})n\right)}.
 \label{eq:spin}
\end{equation}

The small disagreement between the crossing of these analytical curves and 
the beginning of instabilities in the numerical results arises from the 
curvature of the modes near the bifurcation point, and decreases for longer 
rings. This first instability triggers new collisions between modes and, as a 
consequence, more instability regions. The longer the ring the higher the 
number of instability 
regions in the system, approaching the prediction for the infinite case.
It is worth remarking that the bound state mode predicted by Eq. 
(\ref{eq:freq_rel}), 
in excellent agreement with the numerics, does not change with the size of the 
ring.

To analyze the dynamics of dark-dark solitons in the bound state 
allowed by the repulsive interparticle interactions $g_{12}>0$, we
excite the relative motion of the solitons by imposing the initial ansatz
\begin{align}
\begin{aligned}
&\psi_1=\sqrt{n}\tanh\left(z-z_0\right)
\\
&\psi_2=\sqrt{n}\tanh\left(z+z_0\right),
\end{aligned}
\end{align}
and we fix $\mu=1$.
Figures \ref{fig:osci003}--\ref{fig:osci01} show the comparison 
between the subsequent motion of the solitons from the numerical solution of GP 
Eq. \eqref{eq:ebmr_dim}, and the analytical prediction by Eq. 
(\ref{eq:freq_rel})  fitted to 
$y(t)=z_0\cos{wt}$. As can be seen, it provides a reasonable good 
estimate for small $g_{12}$  (Fig. \ref{fig:osci003}) but fails for larger 
$g_{12}$  (Fig. \ref{fig:osci01}) or also long times.

In order to obtain a more quantitative comparison we have  run 
different real time evolutions for varying $g_{12}$ and equal chemical potential 
$\mu=1$. By 
tracking the position of the solitons (at minimum density), we have 
computed their characteristic frequency from a Fourier analysis in time. The 
numerical results are presented in Fig. \ref{fig:fourier}, and show a very 
good agreement with our analytical prediction for small values of 
$g_{12}/g$.

\subsection{Trapped case}

\subsubsection{Attractive interaction between condensates: ${g_{12}<0}$}

\begin{figure}[!b]
	\centering
	\includegraphics[width=\columnwidth]{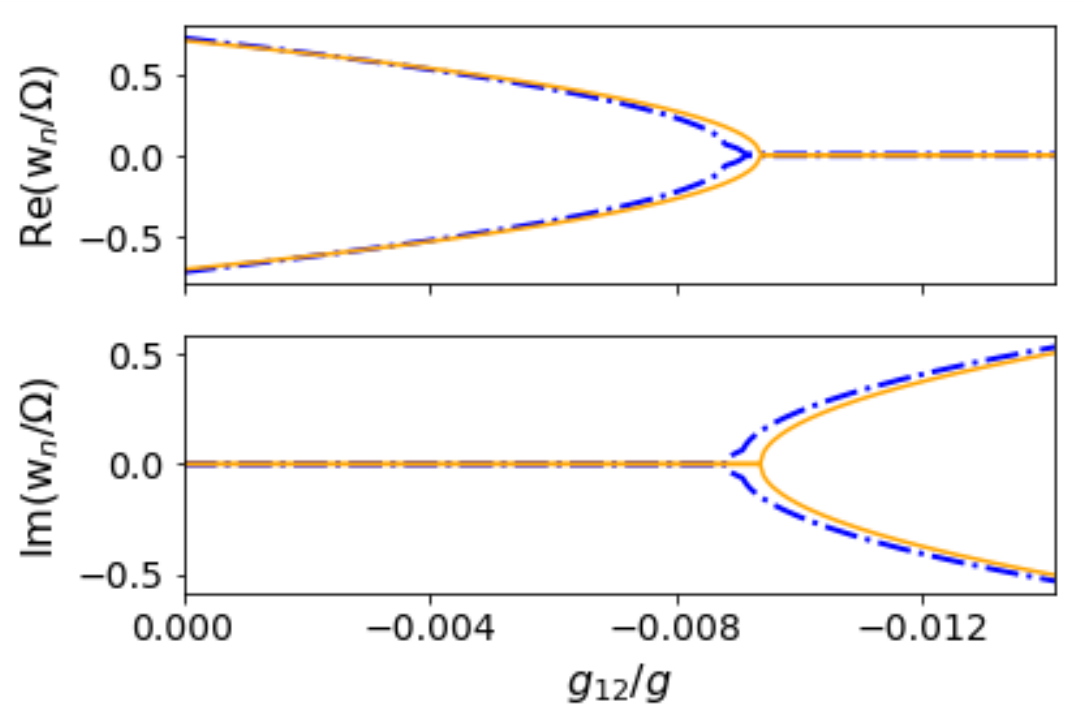}
	\caption{(Color online) Frequency of the dark soliton relative motion in 
a harmonic trap with $\Omega=0.1$. The real (top panel) and 
imaginary (bottom) parts of this frequency are shown according to our 
analytical prediction Eq. 
\eqref{eq:fre} (continuous orange line) and the numerical solution of the 
excitation spectrum from Eq. \eqref{eq:BdG} for the out-of-phase anomalous 
mode (dash-dotted lines).}
	\label{fig:freqg12nega00175}
\end{figure}

As anticipated, in this case the configuration of overlapped 
solitons at $z_0*=0$ is unstable for $|g_{12}/g|>g_c$. This instability can 
also be detected by the appearance of an imaginary frequency in the excitation 
spectrum. Fig. \ref{fig:freqg12nega00175} shows our results for the linear 
excitations of such a stationary state with $\mu=10\, \Omega$ from the 
solution of the Bogoliubov equations (dash-dotted lines). The  frequency of the 
out-of-phase anomalous mode (see a discussion of this mode in next section) 
takes real values for 
$|g_{12}/g|<g_c$ and pure imaginary for $|g_{12}/g|>g_c$. This critical point 
$g_c$ is associated to  the change of the total effective potential, 
from  single-well  to  double-well.

\begin{figure}[!t]
	\centering
	\includegraphics[width=1.\columnwidth]{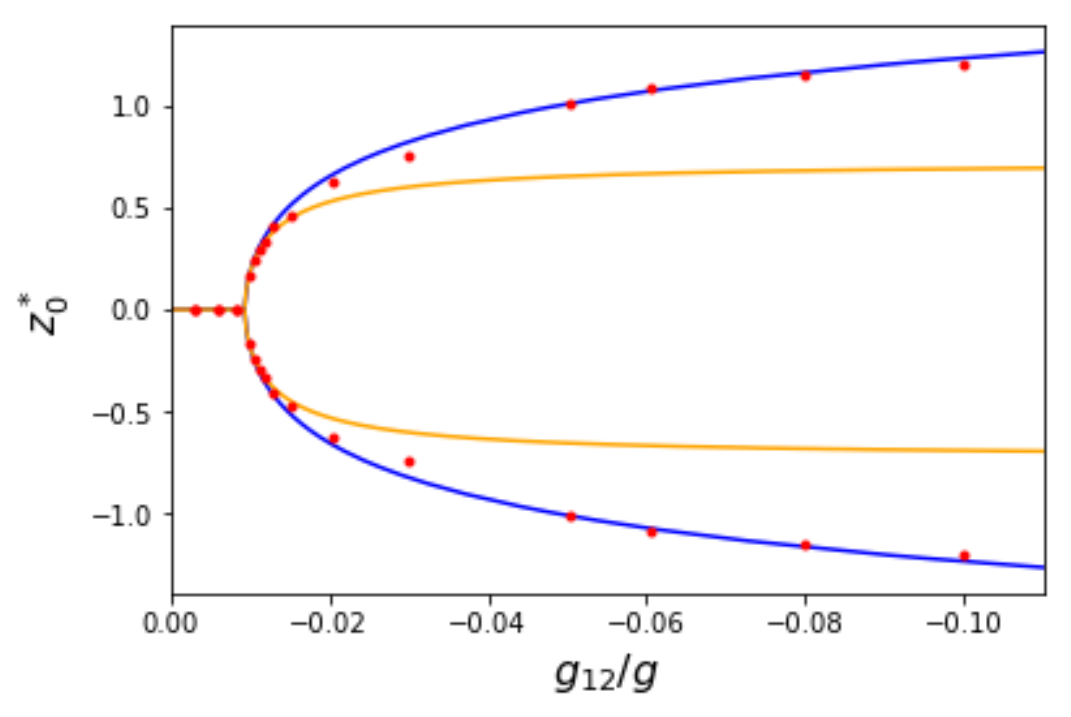}
	\caption{(Color online) Fixed points of the effective double well 
potential given by Eq. \eqref{eq:freq_trap} (continuous blue line)
as a function of $g_{12}/g$. The continuous 
(orange) line represents the analytical result \eqref{eq:fixed_double} and the 
(red) dots are the mean position of the soliton oscillations extracted 
from the numerical solution of GP equation. }
	\label{fig:eqpos2}
\end{figure}

Along with the change of the total effective potential two new stable 
fixed points appear in the system. In Fig. \ref{fig:eqpos2} we compare 
these points (red dots), obtained by extracting the mean position of the 
soliton oscillations around the equilibrium positions,  with 
our analytical approach Eq. \eqref{eq:fixed_double} (orange line).The latter 
fails for increasing values of $|g/g_{12}|$. However the direct 
numerical computation of Eq. \eqref{eq:freq_trap} (blue line) provides a 
very good agreement for the regime of interest, and shows
that the distance between equilibrium points increases with 
$|g_{12}/g|$ instead of being saturated at $2 z_0^*=1.44$.

\begin{figure}[!b]
	\centering
	\includegraphics[width=\columnwidth]{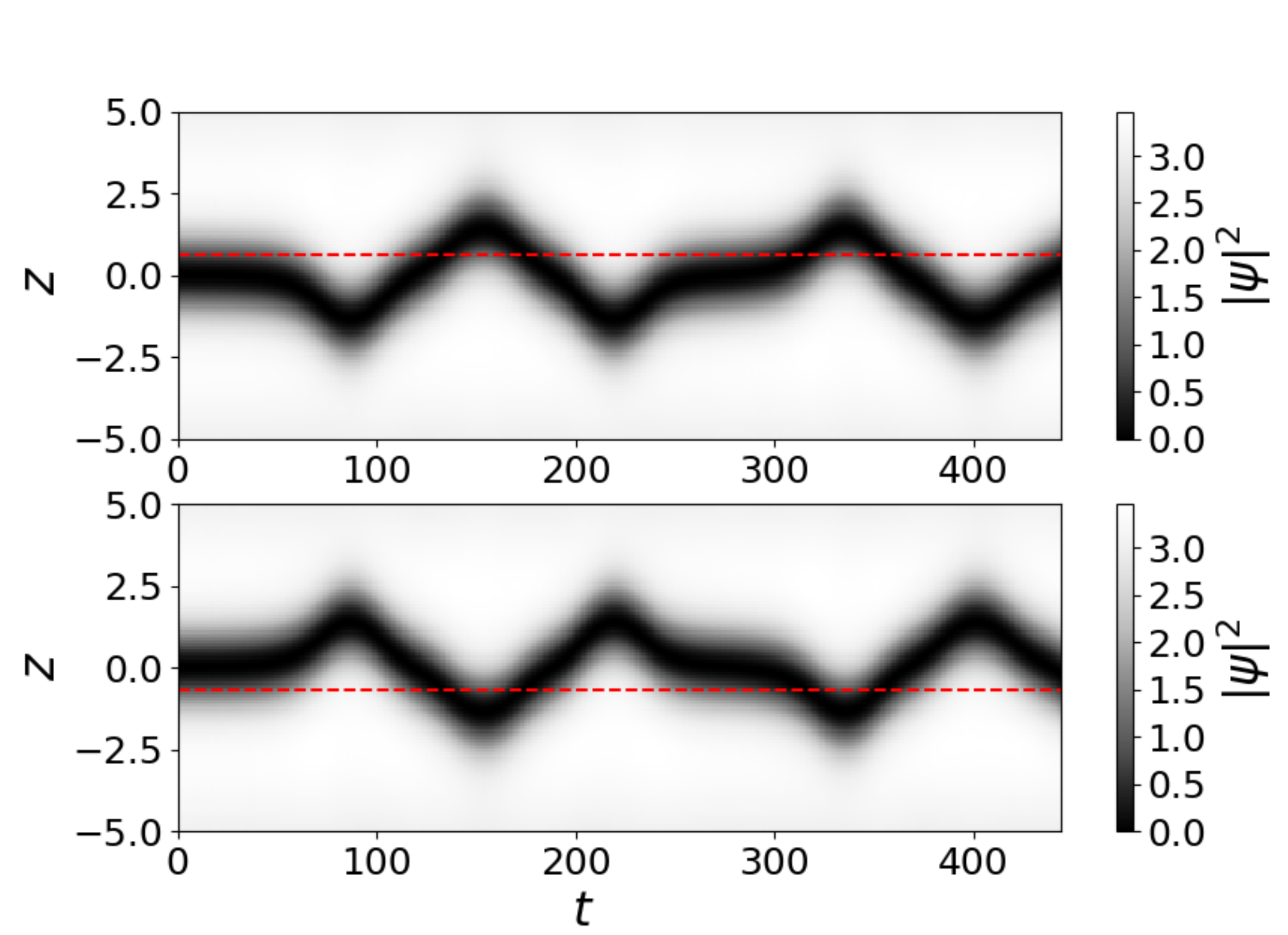}
	\caption{Spatio-temporal evolution of a dark-dark-soliton 
density obtained from the numerical solution of the GP Eq. \eqref{eq:ebmr_dim} 
for $g_{12}/g=-0.046$ and harmonic trapping $\Omega=0.1$. Each 
panel corresponds to a different condensate. The dashed (red) 
lines represent the position of the equilibrium points given by 
\eqref{eq:fixed_double}. 
}
\label{fig:doublewell0046}
\end{figure}

We have numerically solved the GP Eq. 
(\ref{eq:ebmr_dim}) for the real-time evolution of \ dark-dark solitons  
with values of $g_{12}$ 
below and above the bifurcation point $g_c$ for the change of stability. 
First, we have computed a case (see Fig. \ref{fig:doublewell0046}) with 
overlapped solitons situated at $z_0=0$ and $g_{12}/g=-0.046<g_c$, hence 
unstable according to the linear prediction. The initial stationary state has 
been perturbed with white-noise of $1\%$ amplitude. 
As expected the 
system is unstable, and eventually the solitons separate by moving towards the 
minima of the effective double well potential Eq. (\ref{eq:fixed_double}).  
In this case, each dark soliton has enough energy to pass 
through the energy barrier created at the trap center $z_0=0$,  so that 
collisions between them are observed to cause a shift in their trajectories. 
%\hr{Interestingly, two different types of soliton collisions can be 
%observed 
%during the evolution: one that introduces a small shift \cite{Pitaevskii..} 
%(the first collision after separation in Fig. \ref{fig:doublewell0046}, and 
%another that brings the solitons back to a temporary overlap (the second 
%collision in the graph).}

\begin{figure}[!t] 
	\centering
	\includegraphics[width=\columnwidth]{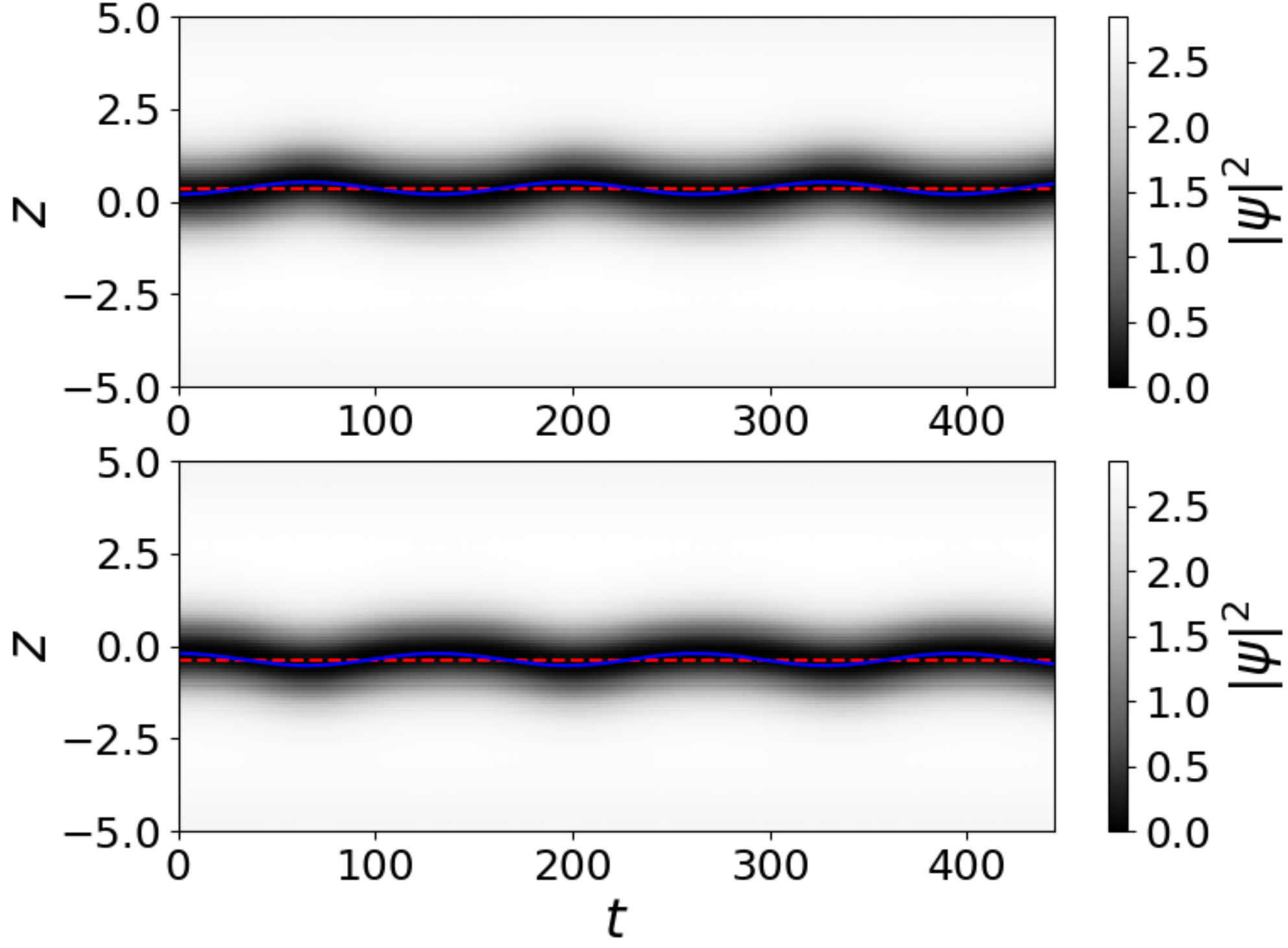}
	\caption{Same as Fig. \ref{fig:doublewell0046} with 
		$g_{12}/g=-0.0125$.
		In this spatio-temporal evolution each dark soliton is situated 
off center ($z_0=\pm0.3$, near the equilibrium 
		points given by \eqref{eq:fixed_double}). The continuous line is 
 fitted according to our analytical results (see text).}
	\label{fig:osci_outcenter_00125}
\end{figure}

If instead the two dark solitons are initially situated at different 
locations, close to the positions of the fixed points Eq. 
(\ref{eq:fixed_double}) (see Fig. \ref{fig:osci_outcenter_00125}), the solitons 
oscillate symmetrically around such points. Their time evolution can be
 fitted by $x(t)=z_0^*+\left(-z_0^*+z_0 \right)\cos(w_{\rm eff}t)$,  where 
$w_{\rm eff}$ is given by \eqref{eq:epsilon}, and provides a good 
approximation to the real time evolution obtained from the GP equation.

\subsubsection{Repulsive interaction between condensates: ${g_{12}>0}$}

\begin{figure}[!t]
	\centering
	\includegraphics[width=1.0\columnwidth]{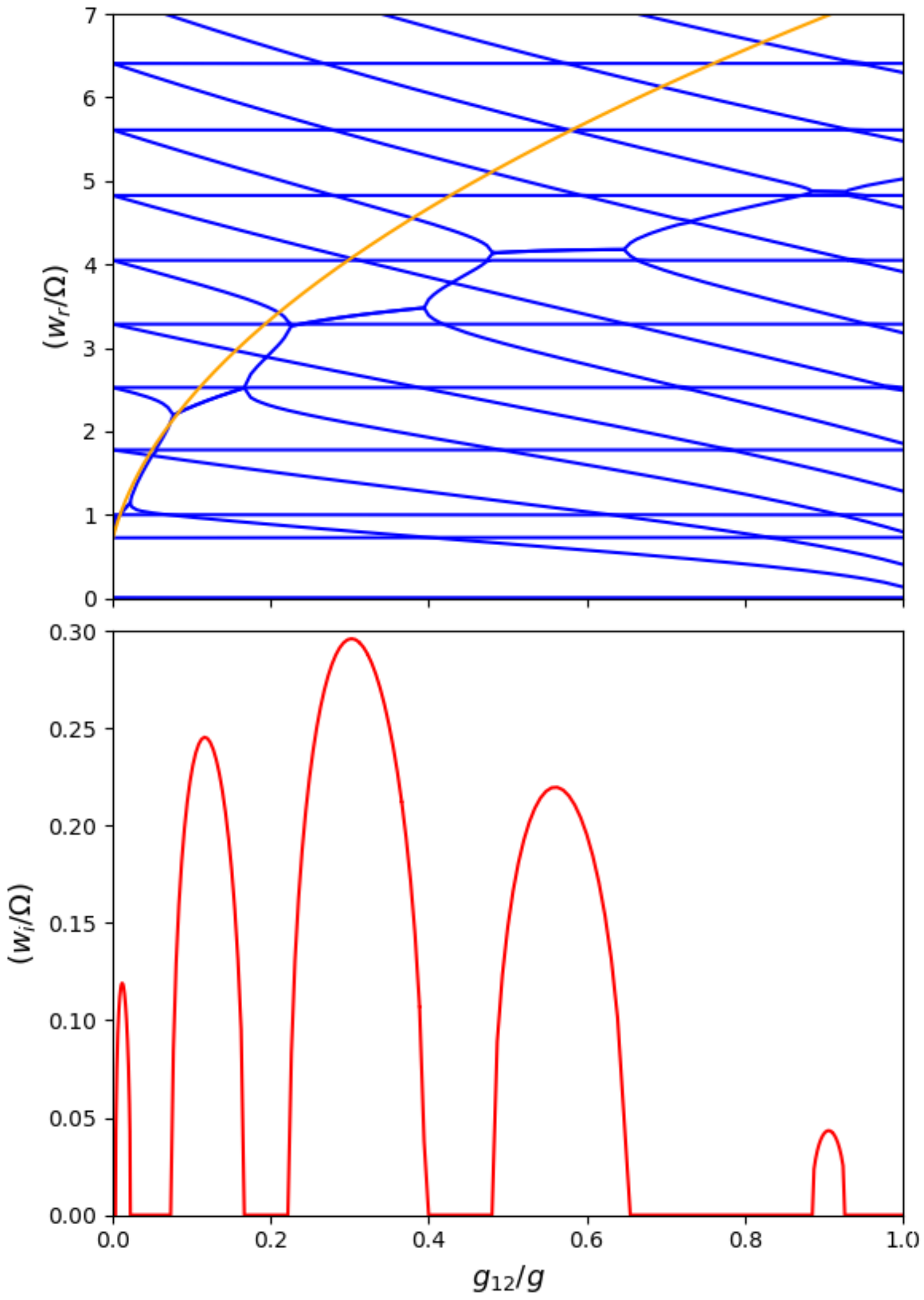}
	\caption{(Color online) Numerical results from the Bogoliubov spectrum 
of a dark-dark soliton with $\mu=10 \,\Omega$ in a trap. The real (top panel) 
and imaginary (bottom panel) parts of the excitation frequencies are shown 
against the interaction strength ratio $g_{12}/g$. 
The continuous (orange) line represents the analytical values from Eq. 
\eqref{eq:freq_rel}.}
	\label{fig:spectrumrealimag}
\end{figure}
Figure \ref{fig:spectrumrealimag} represents the excitation spectrum of 
overlapped  dark-dark solitons at $z_0^*=0$, in the range $g_{12}/g\in[0,1]$, 
for $\mu=10\, \Omega$, within the Thomas Fermi regime of the axial 
harmonic oscillator. At $g_{12}=0$, corresponding to uncoupled solitons, the 
linear modes have double degeneracy, and the lowest energy excitations are 
the anomalous mode, with frequency $\omega_0=\Omega/\sqrt{2}$, and the 
hydro-dynamical excitations $\omega_n=\sqrt{{n(n+1)}/{2}}\,\Omega$, with 
$n=1,2,...$ \cite{Stringari1996,Busch2000,Kevrekidis2015,Frantzeskakis2010}. The 
degeneracy is broken for non null $g_{12}$ and gives rise to two branches of 
in-phase and out-of-phase modes. In the hydro-dynamical case, the in-phase modes 
are associated to excitations in the total density of the background, and 
remain constant for varying $g_{12}$. On the other hand, the out-of-phase
or spin modes account for variations in the difference of the background 
densities \cite{Abad2013}, and decrease their energy for increasing values of 
$g_{12}$.

\begin{figure}[!t]
	\centering
	\includegraphics[width=\columnwidth]{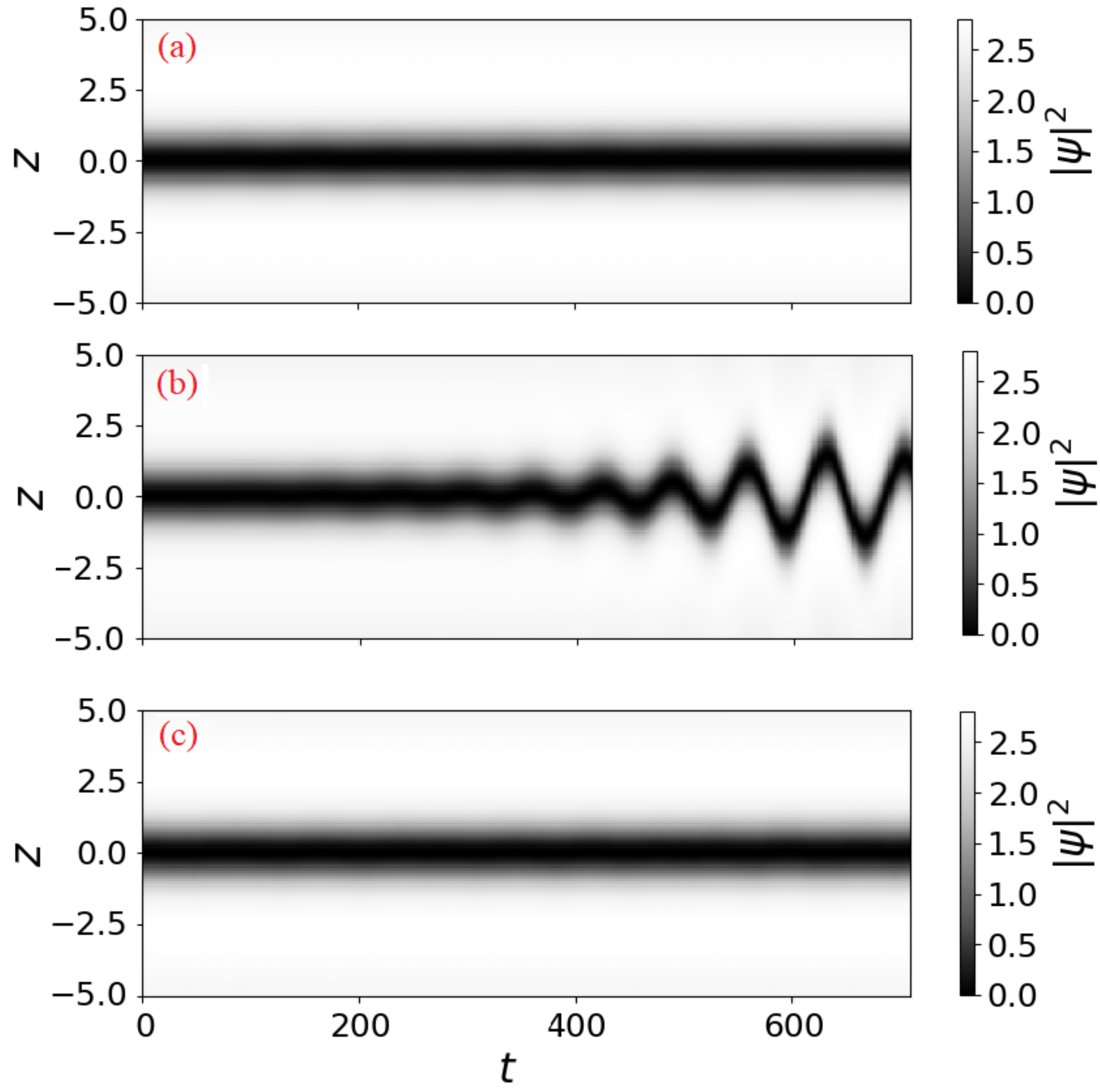}\\
	\includegraphics[width=0.9\columnwidth]{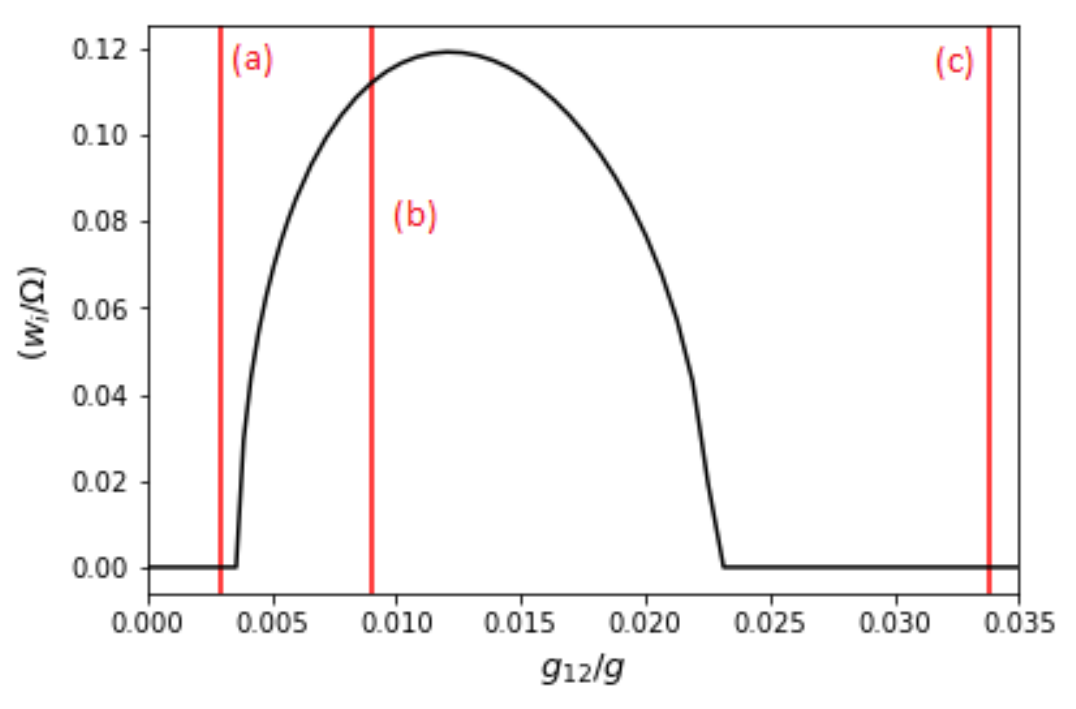}
	\caption{Real time evolution of three  dark-dark 
solitons with varying coupling out of and inside the first instability region 
of Fig. \ref{fig:spectrumrealimag}, showing corresponding stable and unstable 
dynamics. }
	\label{fig:instab_trap}
\end{figure}

The anomalous modes are 
characterized by the negative value of the quantity $norm \times 
energy$ \cite{MacKay1987}, and in scalar condensates
their frequency coincide  with the oscillation frequency 
of the solitons in the trap \cite{Busch2000}.
There are also two different anomalous modes for non null $g_{12}$, 
an in-phase one with constant energy, and 
an out-of-phase mode whose energy increases with $g_{12}$. 
The in-phase anomalous mode is associated with the small amplitude, 
abreast oscillations of the solitons in the trap, hence 
it is the same as in decoupled condensates:
 $\omega_0=\Omega/\sqrt{2}$. However, the out-of-phase 
anomalous mode  is associated with the 
relative motion of the solitons.
This mode, given by the analytical expression Eq. \eqref{eq:fre}, is depicted 
(orange line) in the upper panel of Fig. \ref{fig:spectrumrealimag}, in good 
agreement with the numerical results for small values of $g_{12}/g$.

The lower panel of  Fig. \ref{fig:spectrumrealimag} present our 
numerical results for the imaginary part of the 
spectrum.
As previously commented in the untrapped case, these instabilities are 
characterized by the collision of the out-of-phase 
anomalous mode with the spin mode associated with the background. These 
collisions produce Hamiltonian-Hopf bifurcations where a complex frequency 
quartet appears in the excitation spectrum. It is interesting to see that the 
out-of-phase anomalous mode only collides with odd spin modes. Regions of 
stability and instability alternates up to a value 
of the coupling with $g_{12}/g$ close to 1, but interestingly inside the 
immiscible regime. The higher the chemical 
potential, the closer is this value to $g_{12}/g=1$, according with the 
Bogoliubov analysis for the untrapped case. Such a value can be well 
approximated within the Thomas-Fermi regime by Eq. 
(\ref{eq:spin}) evaluated at maximum density $n=\mu/(g+g_{12})$.
Again the crossing of this analytical frequency with the function for small 
oscillations Eq. (\ref{eq:fre}) provides a good estimate for the 
beginning of the first instability region in the range $g_{12}/g\in[0,1]$.
It is worth noticing that such instability is not captured by Eq. 
\eqref{eq:fre} alone, due to the fact that for its 
derivation the soliton motion was assumed to be decoupled from the 
background.
%However the frequency derived from our theoretical model (Eq. \eqref{eq:fre}) 
%can not predict these instabilities. This is because in our model we have 
%supposed that the motion of the dark solitons is decoupled from the background 
%which is not the case when Hamiltonian-Hopf bifurcations are produced, leading 
%to complex frequency quartets.

Figure \ref{fig:instab_trap} shows examples of stable and unstable 
dark-dark solitons near the first instability region of Fig. 
\ref{fig:spectrumrealimag}. The evolution of the nonlinear systems develops 
according the linear stability analysis and demonstrate the existence of 
dynamically stable coupled solitons, cases (a) and (c), that could be 
experimentally realized.

%\begin{figure}[!tb]
%	\centering
%	\includegraphics[width=\columnwidth]{long_evol_0009.pdf}
%	\caption{Spatio-temporal evolution of an unstable dark-dark soliton 
%obtained from the numerical solution of the GP system 
%\eqref{eq:ebmr_dim} with $\Omega=0.1$ and $g_{12}/g=0.04$. Each panel 
%correspond 
%to a different condensate.}
%	\label{fig:instav1}
%\end{figure}
%To study in more detail the dynamics of the unstable cases, we have numerically
%solve the GP Eq. \ref{eq:ebmr_dim} for two condensates having 
%$\mu=10\,\Omega$ and initially overlapped solitons with $g_{12}/g=0.009$, 
%which corresponds to (b) in Fig.\ref{fig:instab_trap}.
%The initial stationary state has been seeded with 1$\%$-amplitude white-noise. 
%The subsequent evolution is shown in  Fig. \ref{fig:instav1}. It is worth 
%comparing this dynamics with the attractive coupling case of Fig. 
%\ref{fig:doublewell0046}. In the present case, the solitons move in a 
%single-well effective potential, whereas there 
%they move in a effective double well. As a result, in the former case there is 
%no overlapped temporary state after collisions, but still a quasiperiodic 
%behavior alternating three ``weak'' collisions followed by three ``strong'' 
%collisions can be observed.

\section{Conclusions}

The dynamics of dark-dark soliton states in two density coupled BECs has 
been studied within GP theory. By performing a perturbation analysis in the 
parameter $g_{12}/g$ for solitons with equal chemical potential, we have 
derived analytical expressions describing their relative motion both in 
harmonic traps and untrapped systems.
Contrary to the case of solitons in scalar condensates,  our 
theoretical model predicts that the interaction between dark solitons excited 
in different condensates is attractive (repulsive) for repulsive (attractive) 
interparticle interactions $g_{12}$. In harmonically trapped systems, the 
scenario is specially interesting for negative $g_{12}$, where the effective 
potential felt by the solitons modifies its shape as a function of $g_{12}$ 
(through a pitchfork bifurcation) from a single-well potential to a 
double-well potential, then allowing for stationary states made of solitons 
located at different positions.

The theoretical analytical predictions have been shown to be in good agreement with 
the numerical solutions of the Gross Pitaevskii equation for the real 
time evolution, and with the Bogoliubov equations for the linear excitations of 
the dark-dark solitons. In particular, we have demonstrated that the resonance 
of two out-of-phase modes, the anomalous one  giving the frequency of the 
relative motion between solitons, and the lowest energy mode associated to the 
spin density excitation of the background, give rise to instabilities 
(Hopf bifurcations) that produce the decay of dark-dark solitons. This fact
translate in finite systems, either harmonically trapped condensates or ring 
geometries, into alternating regions of dynamical stability and instability. 

The existence of dynamically stable dark-dark solitons open up the way for their 
experimental realization. The current availability of Feshbach resonances for 
tuning both interaction parameters $g$ and $g_{12}$ allows to choose a stable 
fringe in the spectrum. Also in this regard, as a 
natural extension of this work, it would be interesting to explore the stability 
of equivalent states (soliton-soliton or vortex-vortex states) in 
multidimensional systems.

\begin{acknowledgments}
	The authors acknowledge financial support by grants 2014SGR-401 from Generalitat de Catalunya and FIS2014-54672-P from the MINECO (Spain). 
\end{acknowledgments}

\section*{Appendix: Bogoliubov equations for overlapping dark solitons without 
external trap}
In this case, the ground state of Eqs. (\ref{eq:ebmr_dim}) 
is the constant density solution $\psi_1(x)=\psi_2(x)=\sqrt{n}$, such that
$\mu=(g+g_{12})n$. The dark soliton solutions healing to the ground state with 
density $n$ are given by
\begin{equation} 
\psi_{1,2}(x)=\psi_d(x)e^{-i\mu\,t/\hbar}=\sqrt{\frac{\mu}{g+g_{12}}}\, 
\tanh\left(\frac{x}{\xi}\right) e^{-i\mu\,t/\hbar}\,,
 \label{eq:dark}
\end{equation}
where $\xi=\sqrt{\hbar^2/m \mu}$ is the soliton healing length. As in the 
scalar case, it can be seen that the soliton state depends on both, the 
chemical potential $\mu$ and the (sum of the) interaction strength $g+g_{12}$.

We check the stability of Eq. (\ref{eq:dark}) by solving the Bogoliubov 
equations
\begin{align}
 H_0 \, u_1+  \psi_d^2 \left[(2g+g_{12})u_1+g 
v_1 +g_{12}(u_2+v_2) \right] = \omega \, u_1 \nonumber\\
-H_0 \, v_1 -  \psi_d^2 \left[(2g+g_{12})v_1+g 
u_1 +g_{12}(u_2+v_2) \right] = \omega \, v_1 \nonumber\\
 H_0 \, u_2+  \psi_d^2 \left[(2g+g_{12})u_2+g 
v_2 +g_{12}(u_1+v_1) \right] = \omega \, u_2 \nonumber\\
-H_0 \, v_2 -  \psi_d^2 \left[(2g+g_{12})v_2+g 
u_2 +g_{12}(u_1+v_1) \right] = \omega \, v_2 ,
\label{eq:bog0}
\end{align}
where $H_0 = -(\hbar^2/2m)\partial_{xx} -\mu$, and 
$[u_{1}(x), v_{1}(x), u_{2}(x) ,v_{2}(x)]$ are the linear modes with energy 
$\omega$.  

By adding and substracting the the two first previous equations, on the one 
hand, and the two last, on the other hand, we obtain new equations for the 
linear combinations $f_{j\pm}=u_{j}(x) \pm v_{j}(x)$
\begin{align}
 \left(\frac{-\hbar^2}{2m}\partial_{xx}+  \,  (g+g_{12}) \, 
\psi_d^2-\mu\right) f_{1-}= \omega f_{1+} \nonumber\\
\left(\frac{-\hbar^2}{2m}\partial_{xx}+  \, (g+g_{12}) \, 
\psi_d^2-\mu\right) f_{2-}= \omega f_{2+},
\label{eq:bog1-}
\\
 \left( \frac{-\hbar^2}{2m}\partial_{xx} +  (3g+g_{12}) \psi_d^2 \, 
 -\mu \right) f_{1+} +2g_{12}\psi_d^2 f_{2+}=\omega f_{1-} \, \nonumber\\
 \left( \frac{-\hbar^2}{2m}\partial_{xx} +  (3g+g_{12}) \psi_d^2 \, 
 -\mu \right) f_{2+} +2g_{12}\psi_d^2 f_{1+}=\omega f_{2-} \,. 
\label{eq:bog1+}
\end{align}
The first two equations Eqs. (\ref{eq:bog1-}) are already decoupled for the 
modes $1$ and $2$, and (for $\omega=0$) contains zero energy excitations 
associated to the $U(1)$ symmetry presented within each condensate, so that a 
global phase can be arbitrarily picked in the 
soliton solutions Eq. (\ref{eq:dark}). However, Eqs. (\ref{eq:bog1+}) are 
still 
coupling the modes $1$ and $2$. In order to decouple them, we make the 
symmetric $f_{s\pm}=f_{1\pm}(x) + f_{2\pm}(x)$ and antisymmetric 
$f_{a\pm}=f_{1\pm}(x) - f_{2\pm}(x)$ linear combinarions to get
\begin{align}
\left(\frac{-\hbar^2}{2m}\partial_{xx}+  \,  (g+g_{12}) \, 
\psi_d^2-\mu\right) f_{s-}= \omega f_{s+} \nonumber\\
\left( \frac{-\hbar^2}{2m}\partial_{xx} +  3(g+g_{12}) \psi_d^2 \, 
 -\mu \right) f_{s+} =\omega f_{s-} \,,
\label{eq:bog_s}
\end{align}
and
\begin{align}
\left(\frac{-\hbar^2}{2m}\partial_{xx}+  \, (g+g_{12}) \, 
\psi_d^2-\mu\right) f_{a-}= \omega f_{a+} \, \nonumber\\
 \left( \frac{-\hbar^2}{2m}\partial_{xx} +  (3g-g_{12}) \psi_d^2 \, 
 -\mu \right) f_{a+} =\omega f_{a-} \,. 
\label{eq:bog_a}
\end{align}

Equations (\ref{eq:bog_s}) for the symmetric combinations 
$f_{s\pm}=f_{1\pm}(x) + f_{2\pm}(x)$ are equivalent 
to the Bog equations of a scalar condensate in a dark soliton state for an 
interaction strength $g+g_{12}$. As a result they do not present unstable 
modes, and contain two Goldstone modes (with $\omega=0$) reflecting the 
mentioned $U(1)$ symmetry and the translational invariance of the system.

On the other hand, the Eqs. (\ref{eq:bog_a}) for the antisymmetric combinations
$f_{a\pm}=f_{1\pm}(x) - f_{2\pm}(x)$ are relevant for the unstable modes having 
complex frequencies $\omega$.
These modes can first appear at $\omega=0$ for 
particular values of the system parameters $\{\mu, g, g_{12}\}$ (for a given 
particle species of mass $m$), indicating a bifurcation of the soliton 
solutions giving place to a new stationary state. However, as we have seen 
in the text, the instabilities can also appear from a couple of nonzero real 
frequencies which become complex, indicating a Hopf bifurcation with a 
subsequent oscillatory dynamics. Below, we analyze the first case 
associated to the existence of zero modes.

At $\omega=0$, since the first 
equation (\ref{eq:bog_s}) provides the commented Goldstone 
modes, we only have to look for solutions to
\begin{align}
 \left( \frac{-\hbar^2}{2m}\partial_{xx} +  \mu 
\frac{3g-g_{12}}{g+g_{12}}\, 
\tanh^2\left(\frac{x}{\xi} \, 
\right)-\mu \right)f_{a+} =0\,,
\end{align}
that in units of the healing length gives 
\begin{align}
 -\frac{1}{2}\partial_{xx} f_{a+}- (1+\delta g) \, 
\mbox{sech}^2(x) f_{a+} = -\delta g \, f_{a+},
\label{eq:zero}
\end{align}
where $\delta g=2(g-g_{12})/({g+g_{12}})=2(g-g_{12})n/\mu$ is the 
relevant parameter for the emergence of instabilities. Eq. (\ref{eq:zero}) is a 
Schr\"odinger equation for the asymmetric wavefunction $f_{a+}$ in the potential 
well $\mbox{sech}^2(x)$ with depth $(1+\delta g)$. This well has bound states 
with energy $-\delta g$ whenever \cite{Rosen, Sophie2017}
\begin{equation}
\sqrt{2 \,\delta g+\frac{9}{4}}-\sqrt{2\,\delta g}-\frac{1}{2}=n \, ,
\end{equation}
\linebreak
with $n=0,1,2,\dots$ and $n \leq \sqrt{2\delta 
g+9/4}-1/2$. The latter condition ensures that $g_{12}<g$ and saturates with 
$g=g_{12}$ at the starting point of the inmiscible regime.

\end{document}